\documentclass[a4paper]{jpconf}
\usepackage{graphicx}
\begin{document}
\title{NA61/SHINE ion program}

\author{Maja Mackowiak for the NA61 Collaboration}

\address{Warsaw University of Technology, Faculty of Physics, Koszykowa 75, 00-662 Warszawa, Poland}

\ead{majam@if.pw.edu.pl}

\begin{abstract}
The Super Proton Synchrotron (SPS) at CERN covers one of the most interesting regions of the phase diagram (T - $\mu_{B}$) of strongly interacting matter. The study of central Pb+Pb collisions by NA49 indicate that the threshold for deconfinement is reached already at the low SPS energies. Theoretical considerations predict a critical point of strongly interacting matter at energies accessible at the SPS. The NA61/SHINE experiment, a successor of the NA49 project, will study hadron production in p+p, p+A, h+A, and A+A reactions at various energies. The broad physics program includes the investigation of the properties of strongly interacting matter, as well as precision measurements of hadron spectra for the T2K neutrino experiment and for the Pierre Auger Observatory and KASCADE cosmic-ray projects. The main physics goals of the NA61/SHINE ion program are to study the properties of the onset of deconfinement at low SPS energies and to find signatures of the critical point of strongly interacting matter. To achieve these goals a broad range in the (T - $\mu_{B}$) phase diagram will be covered by performing an energy ($10A$-$158A$ GeV/c) and system size (p+p, B+C, Ar+Ca, Xe+La) scan. The first data for this 2-D scan were taken in 2009, i.e. p+p interactions at $20$, $30$, $40$, $80$, $158$ GeV/c beam energy. This contribution will summarize physics arguments for the NA61/SHINE ion program, show the detector performance and present the current status of the experiment and plans for the next years.
\end{abstract}

\vspace{0.5cm}
\noindent{\bf The NA61 Collaboration:}
\noindent
N.~Abgrall${}^{22}$,
A.~Aduszkiewicz${}^{23}$,
B.~Andrieu${}^{11}$,
T.~Anticic${}^{13}$,
N.~Antoniou${}^{18}$,
J.~Argyriades${}^{22}$,
A.~G.~Asryan${}^{15}$,
B.~Baatar${}^{9}$,
A.~Blondel${}^{22}$,
J.~Blumer${}^{5}$
% ...

\vspace*{1cm}

\noindent
${}^{ 1}$The Universidad Tecnica Federico Santa Maria, Valparaiso, Chile  \\
${}^{ 2}$ETH, Zurich, Switzerland \\
${}^{ 3}$Fachhochschule Frankfurt, Frankfurt, Germany \\
${}^{ 4}$Faculty of Physics, University of Sofia, Sofia, Bulgaria \\
${}^{ 5}$Forschungszentrum Karlsruhe, Karlsruhe, Germany \\
${}^{ 6}$Institute for Nuclear Research, Moscow, Russia \\
${}^{ 7}$Institute for Particle and Nuclear Studies, KEK, Tsukuba,  Japan \\
${}^{ 8}$Jagiellonian University, Cracow, Poland  \\
${}^{ 9}$Joint Institute for Nuclear Research, Dubna, Russia \\
${}^{10}$KFKI Research Institute for Particle and Nuclear Physics, Budapest, Hungary \\
${}^{11}$LPNHE, University of Paris VI and VII, Paris, France \\
${}^{12}$University of Silesia, Katowice, Poland  \\
${}^{13}$Rudjer Boskovic Institute, Zagreb, Croatia \\
${}^{14}$Soltan Institute for Nuclear Studies, Warsaw, Poland \\
${}^{15}$St. Petersburg State University, St. Petersburg, Russia \\
${}^{16}$State University of New York, Stony Brook, USA \\
${}^{17}$Jan Kochanowski University in  Kielce, Poland \\
${}^{18}$University of Athens, Athens, Greece \\
${}^{19}$University of Bergen, Bergen, Norway \\
${}^{20}$University of Bern, Bern, Switzerland \\
${}^{21}$University of Frankfurt, Frankfurt, Germany \\
${}^{22}$University of Geneva, Geneva, Switzerland \\
${}^{23}$University of Warsaw, Warsaw, Poland \\
${}^{24}$Warsaw University of Technology, Warsaw, Poland  \\
${}^{*}$ {\it deceased}  \\

\section{Introduction}
The NA61/SHINE experiment (Fig. \ref{layout}) \cite{proposal} is located at the SPS accelerator at CERN. It is the continuation of the fixed target NA49 experiment and will study hadron+hadron, hadron+nucleus and nucleus+nucleus collisions. The NA61/SHINE collaboration consists of 130 physicists from 24 institutions. SHINE stands for {\it SPS Heavy Ion and Neutrino Experiment}. The NA61/SHINE physics program includes:
\begin{itemize}
\item hadron production measurements for neutrino and cosmic-ray experiments (not discussed here (for details see \cite{proposal})
\item ion collision program studying the onset of deconfinement, searching for the critical point of strongly interacting matter and investigating high $p_{T}$ physics
\end{itemize} 

\section{Detector}
\begin{figure}[h]
\begin{minipage}{18pc}
\includegraphics[width=18pc]{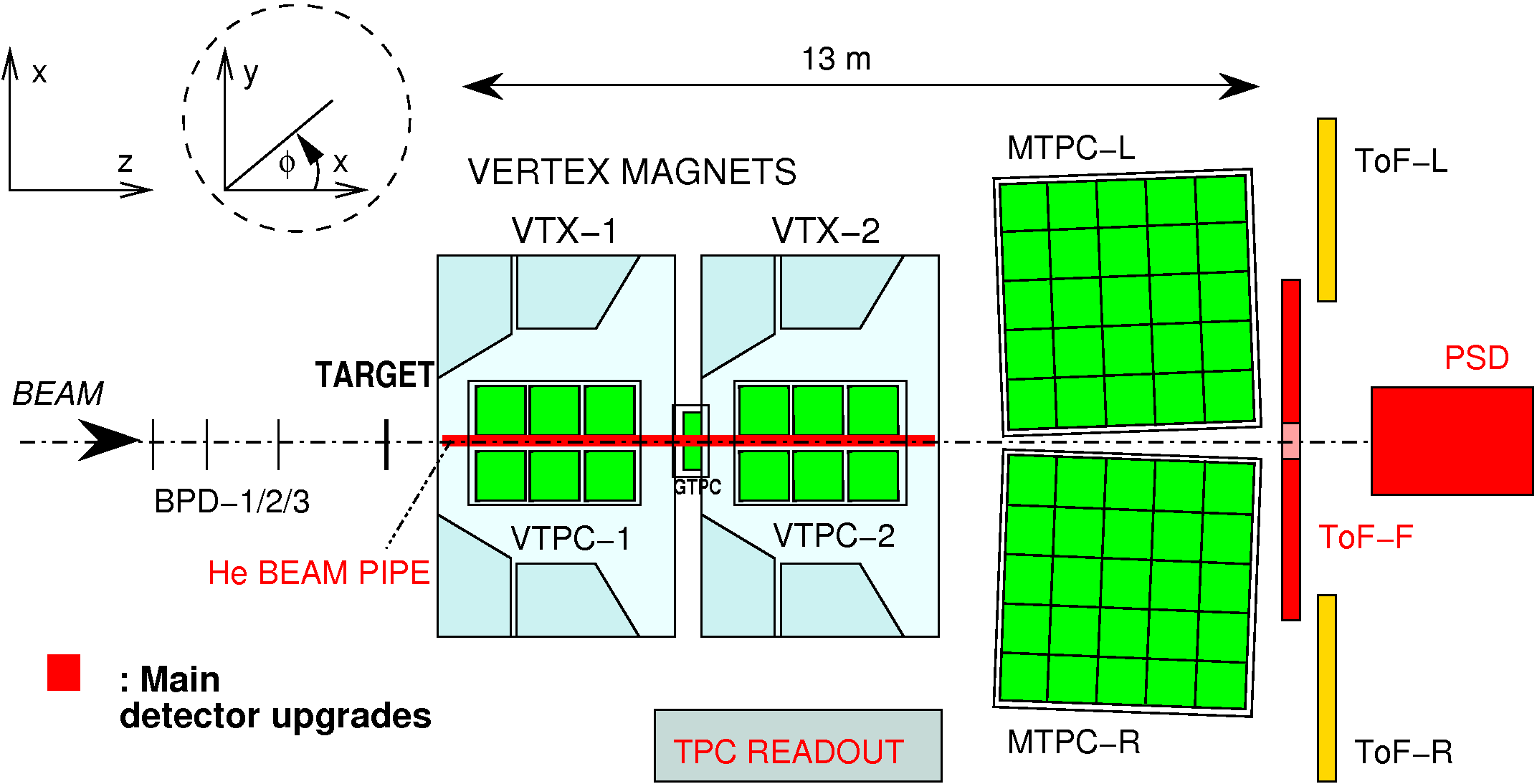}
\newline
\caption{\label{layout} \small NA61/SHINE detector layout. Upgrades are shown in red color.}
\end{minipage}\hspace{2pc}%
\begin{minipage}{18pc}

\includegraphics[width=18pc]{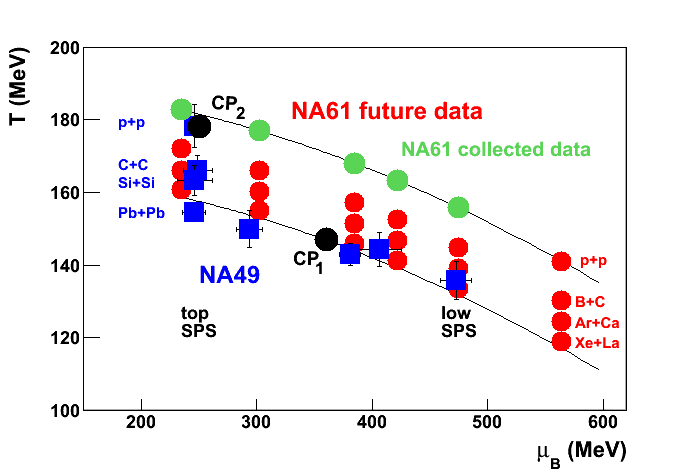}
\vspace{-2pc}
\caption{\label{2Dscan} \small System size and energy scan of the phase diagram. Estimated (squares [NA49]) and extrapolated (circles) chemical freeze-out points (\cite{becatini}). Green circles indicate data collected in 2009. Black circles - two locations of critical point analyzed by NA49.}
\end{minipage} 
\end{figure}

The main component of the NA61/SHINE detector (Fig. \ref{layout}) are four large volume Time Projections Chambers. The first two (VTPC-1/2) are located in the magnetic field of two superconducting dipole magnets. The other two (MTPC-L/R) are positioned downstream of the magnets. An additional small gap TPC (GTPC) is installed on the beam axis between VTPC-1 and VTPC-2. Behind the TPCs are three ToF-walls. Two of them (ToF-L/R) were inherited from the NA49 experiment. A forward ToF-wall was added in 2007 to extend particle identification in the low momentum range. \newline
In 2008 a new TPC read-out system was constructed which increased the event recording rate by a factor of 10 compared to NA49. Two other upgrades are planned mainly for the ion beam program. The first one (partially installed in 2010) is the Projectile Spectator Detector (PSD), which will allow precise measurement of the energy of projectile spectators and a reconstruction of the reaction plane. The determination of the number of projectile spectators is crucial for measurements of event-by-event multiplicity fluctuations. The PSD will allow single nucleon discrimination. The second upgrade is a He-beam pipe which will reduce $\delta$-electron production by the beam by a factor of 10. It will be installed in 2011.

\section{NA61/SHINE ion program}
One of the most important tasks of heavy ion collision physics is to explore the phase diagram of strongly interacting matter. Theorists expect that for large values of baryon chemical potential ($\mu_{B}$) there is a first order phase transition between hadron matter and quark gluon plasma. In contrast, for small $\mu_{B}$ values the transition is expected to be a smooth crossover. The $1^{st}$-order phase transition line ends with a critical point of a $2^{nd}$-order phase transition. One of the main goals of NA61/SHINE is to study the transition line and search for the critical point. To achieve these goals NA61/SHINE will perform a 2-D scan (energy and system size) of the phase diagram (Fig. \ref{2Dscan}). The first part of the scan, i.e. collection of p+p collisions was/is done in 2009/2010. Also in 2010 a test of a B beam from fragmentation of Pb ions will be performed. B+C interactions will be collected in 2011(13). Ar+Ca and Xe+La are planned for 2012 and 2014. p+Pb interactions will be recorded in 2011/2012.  
\begin{figure}[h]
\begin{minipage}{14pc}
\vspace{-4pc}
\includegraphics[width=14pc]{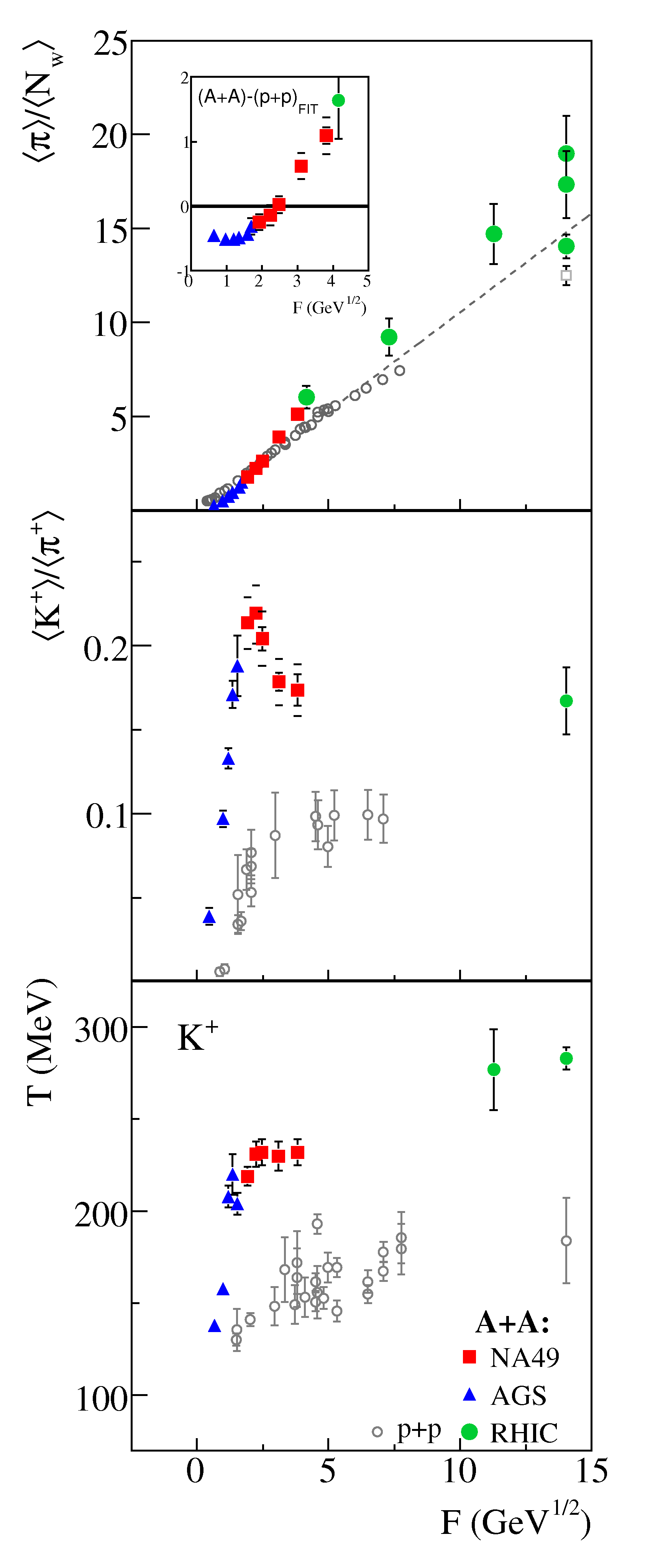}
\caption{\label{ood} \small Signals of the onset of deconfinement: kink, horn, step \cite{ood}. Measurements are plotted as a function of the Fermi measure $F\approx\sqrt{\sqrt{s_{NN}}}$.}
\end{minipage}\hspace{2pc}%
\begin{minipage}{22pc}
\vspace{-6pc}
\includegraphics[width=22pc]{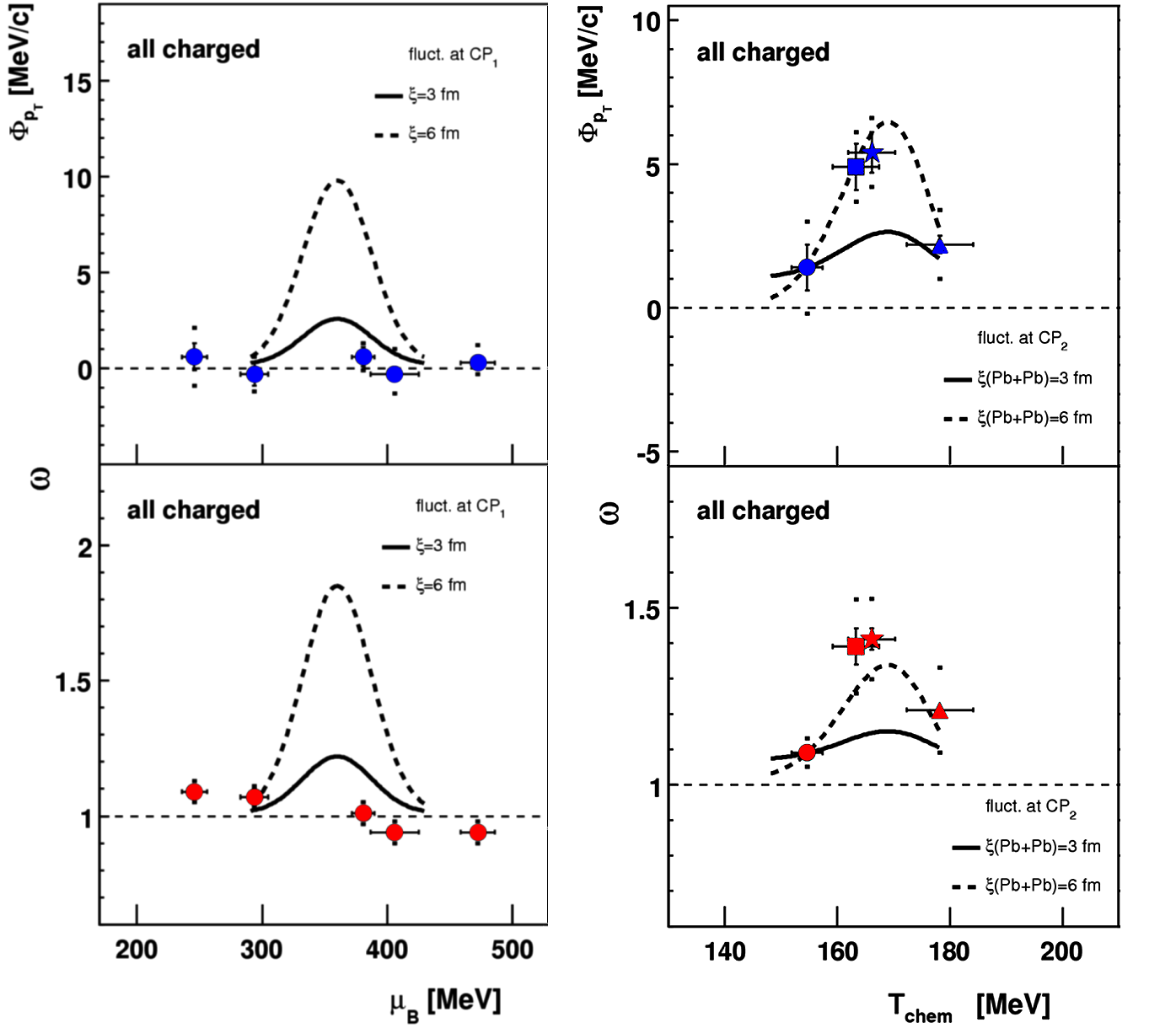}
\caption{\label{CP12} \small Transverse momentum and multiplicity fluctuations measured by NA49. Left: energy scan for central Pb+Pb collisions. Right: system size dependence (p+p, C+C, Si+Si and Pb+Pb) at 158A GeV. Lines indicate predictions for the critical point calculated for its two locations with two values of the correlation length parameter (dashed and solid lines). For details see Ref. \cite{kasiaQM, mazury}, and references therein.}
\end{minipage} 
\end{figure}
\subsection{Onset of deconfinement}
When the early stage of matter created in an ion collision hits the transition line three signals should be observed (so-called: kink, horn, step) according to the SMES model \cite{smes}. Such signals have been observed in the NA49 experiment in central Pb+Pb collisions near $30A$ GeV energy \cite{ood} (Fig. \ref{ood}). In contrast to the structures seen in Pb+Pb (Au+Au) reactions a smooth behavior is seen for elementary interactions. NA61/SHINE will check where the evidence of the onset of deconfinement appears for light systems (B+C, Ar+Ca, Xe+La).
\subsection{Critical Point}
The critical point (CP) of the phase transition is predicted to lie in a region of the phase diagram accessible at SPS energies \cite{katz}. Enhanced fluctuations in multiplicity and transverse momentum are expected if hadronization and freeze-out happen near the critical point \cite{stephanov}. The NA49 experiment has already looked for indications of the critical point by studying average $p_{T}$ ($\phi_{p_{T}}$ measure) and multiplicity fluctuations ($\omega$) (\cite{kasiaQM, mazury}, and references therein). NA49 considered two locations of the CP (shown in Fig. \ref{CP12}). The second location of the CP ($\mu_{B}$ for A+A collisions at $158A$ GeV, $T_{chem}$ for p+p collisions at $158$ GeV) is consistent with the data. NA61/SHINE will search for a maximum of fluctuation by a 2-D scan (Fig. \ref{2Dscan}). 
\subsection{High $p_{T}$ physics} 
\begin{figure}[h]
\center

\includegraphics[width=26pc]{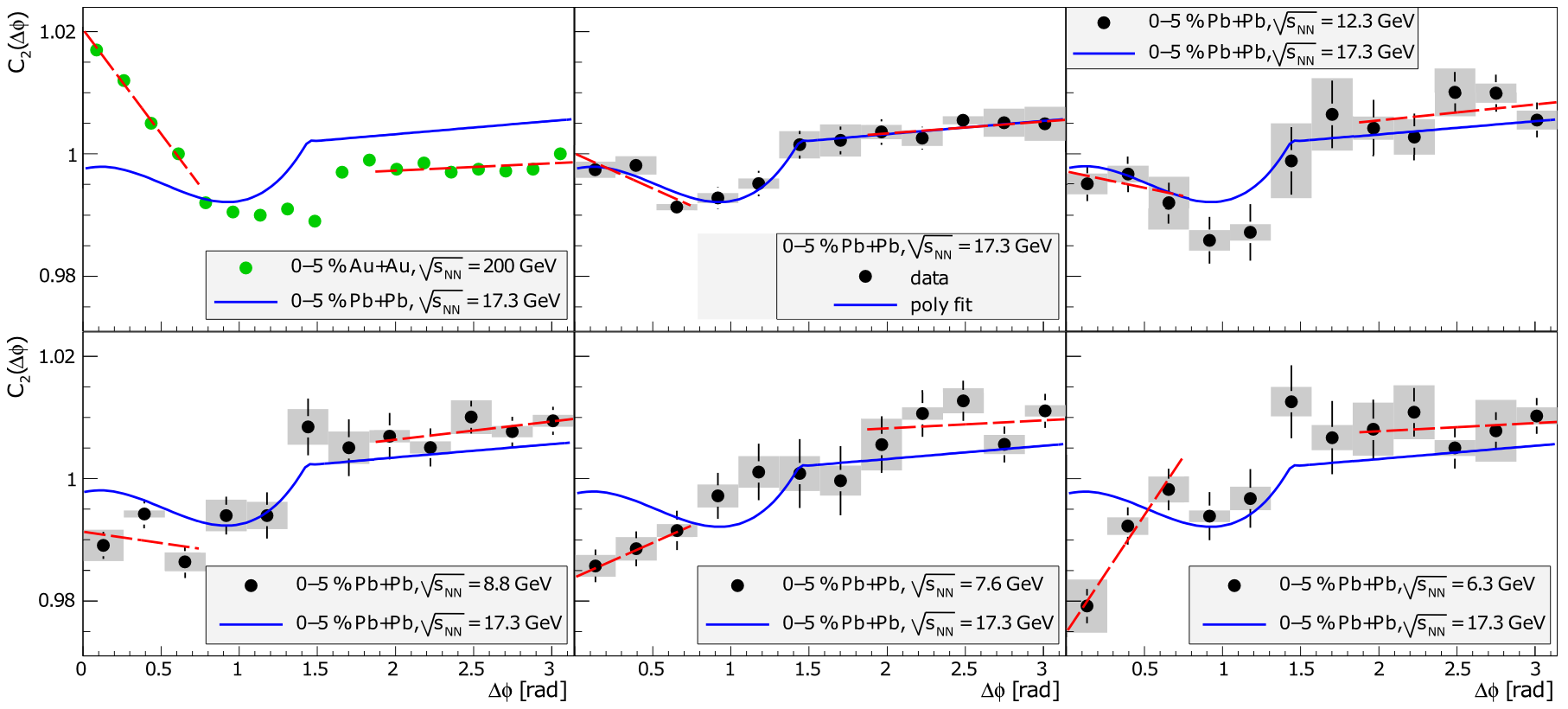}
\caption{\label{marek} \small Two-particle azimuthal angle correlation function for central Pb+Pb (Au+Au) events at $\sqrt{s_{NN}}$= 200, 17.3, 12.3, 8.8, 7.6 and 6.3 GeV (full symbols) compared to results for central Pb+Pb collisions at 17.3 GeV. Curves represent the correlation function at $\sqrt{s_{NN}}$= 17.3 GeV and is shown for comparison at the other energies (for details see \cite{QM2008, QM2009}).}
\end{figure}

There are two very interesting phenomena in high $p_{T}$ physics which will be studied in NA61/SHINE. The first one is suppression of high $p_{T}$ particles observed by RHIC experiments at $\sqrt{s_{NN}}=62-200$ GeV collision energy \cite{rhic}. It would be interesting to check whether suppression appears already at SPS energies. However, there is not sufficient statistics for p+p and p+Pb collisions in the NA49 experiment. NA61/SHINE will increase these statistics by an order of magnitude. Moreover, the system size and energy dependence of the nuclear modification factor $R_{AA}$ will be investigated (for details see \cite{andreas}). \newline
The second interesting phenomenon, presented in \cite{QM2008,QM2009}, is a change of the shape of the near-side peak in the two-particle azimuthal angle correlation function with decreasing collision energy in the SPS regime. It is the so-called Jet-Hole transition presented in Fig. \ref{marek}. 
The origin of this phenomenon is not known yet and further studies are required. 
\vspace{-1pc}
\section{Summary}
The NA61/SHINE program gives the unique opportunity to reach exciting physics in a very efficient and cost effective way. There are several other complementary projects such as the RHIC low energy scan (starting in 2010), NICA (starting in 2014) and SIS-100(SIS-300) (starting in 2015/17). The advantages of the NA61/SHINE ion program over the RHIC energy scan are:
\begin{itemize}
\item 2-D scan (energy and system size) of the phase diagram
\item measurements of identified hadron spectra in a broad rapidity range, which in particular allows to obtain mean hadron multiplicities in full phase space
\item measurements of the total number of projectile spectators including free nucleons and nucleons in nuclear fragments
\item high event rate in the full SPS energy range including the lowest energies
\item low $p_{T}$ region accessible (signatures of the critical point should be visible mostly in the low $p_{T}$ region)
\end{itemize}     
On the other hand, the advantage of collider experiments is the complete azimuthal acceptance which does not depend on energy (canceling of many systematic uncertainties).

\ack
{This work was supported by  
the Hungarian Scientific Research Fund (OTKA 68506),
the Polish Ministry of Science and Higher Education (N N202 3956 33),
the Federal Agency of Education of the Ministry of Education and Science
of the Russian Federation (grant RNP 2.2.2.2.1547) and
the Russian Foundation for Basic Research (grants 08-02-00018 and 09-02-00664),
the Ministry of Education, Culture, Sports, Science and Technology,
Japan, Grant-in-Aid for Scientific Research (18071005, 19034011,
19740162),
Swiss Nationalfonds Foundation 200020-117913/1 
and ETH Research Grant TH-01 07-3.
}

\end{document}